\documentclass[amsmath,tightenlines,superscriptaddress,twocolumn,amssymb,prb,aps]{revtex4}

\usepackage{graphicx}
\usepackage{dcolumn}
\usepackage{pdfsync}
\usepackage{bm}
\usepackage{array}
\usepackage{hyperref}

\begin{document}
\title{Predicted alternative structure for tantalum metal under high pressure and high temperature}
\author{Zhong-Li Liu}
\email{zl.liu@163.com}

\affiliation{College of Physics and Electric Information, Luoyang Normal University, Luoyang 471022, China}
\affiliation{Laboratory for Shock Wave and Detonation Physics Research, Institute of Fluid Physics, P.O. Box 919-102, 621900 Mianyang, Sichuan, China}

\author{Ling-Cang Cai}

\affiliation{Laboratory for Shock Wave and Detonation Physics Research, Institute of Fluid Physics, P.O. Box 919-102, 621900 Mianyang, Sichuan, China}

\author{Xiu-Lu Zhang}
\affiliation{Laboratory for Shock Wave and Detonation Physics Research, Institute of Fluid Physics, P.O. Box 919-102, 621900 Mianyang, Sichuan, China}
\affiliation{Laboratory for Extreme Conditions Matter Properties, Southwest University of Science and Technology, 621010 Mianyang, Sichuan, China}

\author{Feng Xi}
\affiliation{Laboratory for Shock Wave and Detonation Physics Research, Institute of Fluid Physics, P.O. Box 919-102, 621900 Mianyang, Sichuan, China}

\date{\today}

\begin{abstract}
First-principles simulations have been performed to investigate the phase stability of tantalum metal under high pressure and high temperature (HPHT). We searched its low-energy structures globally using our developed multi-algorithm collaborative (MAC) crystal structure prediction technique. The body-centred cubic (bcc) was found to be stable at pressure up to 300 GPa. The previously reported $\omega$ and \textit{A}15 structures were also reproduced successfully. More interestingly, we observed another phase (space group: \textit{Pnma}, 62) that is more stable than $\omega$ and \textit{A}15. Its stability is confirmed by its phonon spectra and elastic constants. For $\omega$-Ta, the calculated elastic constants and high-temperature phonon spectra both imply that it is neither mechanically nor dynamically stable. Thus, $\omega$ is not the structure to which bcc-Ta transits before melting. On the contrary, the good agreement of \textit{Pnma}-Ta shear sound velocities with experiment suggests \textit{Pnma} is the new structure of Ta implied by the discontinuation of shear sound velocities in recent shock experiment [J. Appl. Phys. \textbf{111},  033511 (2012)].
\end{abstract}
  
\maketitle

\section{Introduction}
Tantalum (Ta) metal is used frequently as the pressure scale in the diamond-anvil cell (DAC) and shock wave (SW) experiments, thanks to its high stability, chemical inertness, and very high melting temperature. It has attracted tremendous interests on its very wide range of properties in recent years. Among these properties, melting is the most intriguing one because there are very large discrepancies in its high-pressure melting temperature between different experiments. In detail, there exist several thousand degrees of discrepancies in the melting temperature of Ta when extrapolating from the diamond-anvil cell (DAC)~\cite{Errandonea2001,Errandonea2003} pressures of $\sim$100 GPa to the shock wave (SW)~\cite{Brown1984} pressure of $\sim$300 GPa. Extensive investigations have been performed to understand the underlying causes both experimentally and theoretically.~\cite{Foata-Prestavoine2007,Taioli2007,Liu2008,Errandonea2005,Luo2007,Ross2007-1,Verma2004,Wang2001,Moriarty2002,Wu2009,Burakovsky2010,Dewaele2010,Klug2010,Haskins2012,Ruiz-Fuertesa2010} Nevertheless, the melting discrepancies of Ta still remain inconclusive up to now.

Most recently, Dewaele \textit{et al.}~\cite{Dewaele2010} revisited the melting curve of Ta via DAC experiment and obtained a much higher result compared to previously reported flat DAC melting curve~\cite{Errandonea2001,Errandonea2003}. In their DAC experiment, they excluded the effects of chemical reactivity of Ta samples with pressure medium and pressure medium melting. And they believe that the previous tantalum melting curve~\cite{Errandonea2001,Errandonea2003} has been underestimated because of the undetected chemical reactions or errors of pyrometry. Furthermore, the measured melting data are in general agreement with our previous molecular dynamics simulations~\cite{Liu2008} and \textit{ab initio} results by Taioli \textit{et al.}~\cite{Taioli2007} with the difference of around 1000 K. While, almost at the same time, in another new DAC experiment Ruiz-Fuertesa \textit{et al.} still obtained the same flat melting curve,~\cite{Ruiz-Fuertesa2010} in contrast with the DAC results of Dewaele \textit{et al.}~\cite{Dewaele2010}

Whether or not a solid-solid (SS) phase transition occurs before melting is key to understand the large discrepancies between DAC and SW experiments, and those between different DAC experiments. So, the SS transition in Ta has been another attractive subject in recent years. Theoretically, Burakovsky \textit{et al.} reported a phase transition from bcc to hexagonal omega (hex-$\omega$) phase above 70 GPa based on first-principles simulations.~\cite{Burakovsky2010} They also believe that the previous DAC melting data~\cite{Errandonea2001,Errandonea2003} are problematic, because the observed sample motion in DAC experiments is very likely not due to melting but internal stresses accompanying a SS transition such as bcc-$\omega$ transition.~\cite{Burakovsky2010} In addition, in highly undercooled tantalum liquid Jakse \textit{et al.} observed another metastable structure, \textit{A}15, in the molecular dynamics simulations.~\cite{Jakse2004} While, no SS phase transition was observed in the new DAC experiment by Dewaele \textit{et al.}~\cite{Dewaele2010}

On one hand, there is at least about 1000 K regime between the new DAC data~\cite{Dewaele2010} and the theoretical predictions,~\cite{Liu2008,Taioli2007} where it is possible to occur SS phase transitions. On the other hand, the discontinuity in recent experimental shear sound velocities $C_l$ at $\sim$60 GPa~\cite{Hu2012} and the softening of experimental and theoretical shear sound velocities~\cite{Antonangeli2010} have the obvious characteristics of SS phase transitions. Furthermore, Hsiung observed the displacive $\omega$ phase transition within the shock-recovered polycrystalline tantalum after being shocked to relatively lower pressure, 45 GPa.~\cite{Hsiung1998-1,Hsiung2000,Hsiung2010} It was recently reported that the observed $\omega$ phase in the shock-recovered Ta is not ideal hexagonal but pseudo-hexagonal.~\cite{Hsiung2010} But what Burakovsky~\textit{et al.}~\cite{Burakovsky2010} reported is the ideal $\omega$ phase. So, it is necessary to further explore the phase stability of Ta at high pressure and high temperature (HPHT) to understand the high-pressure melting discrepancies between different experiments. This motivates us to further investigate whether or not bcc-Ta transit to other phase before melting.

First, we globally searched the low-energy structures of Ta using our recently developed \textit{ab initio} MAC crystal structure prediction technique. Then, the stability of the produced metastable structures was carefully checked. Finally, the sound velocities of the stable and metastable structures were deduced from the calculated elastic constants and compared with experiments.

The rest of this paper is organized as follows. Section~\ref{compdet} is the computational details. We present the results and discussion in Section~\ref{results}. The conclusions are drawn in Section~\ref{concl}.

\section{Computational details}
\label{compdet}

In order to determine the stable structures and alternative metastable structures for Ta, we searched its low-energy structures from 0 to 300 GPa with the interval of 10 GPa using the MAC crystal structure prediction technique as implemented in our {\sc Muse} code. The MAC algorithm combines organically the multi algorithms including the evolutionary algorithm, the simulated annealing algorithm and the basin hopping algorithm to search collaboratively the global energy minima of materials with the fixed stoichiometry.~\cite{muse} After introduced the competition in all the evolutionary and variation operators, the evolution of the crystal population and the choice of the operators are self-adaptive automatically. That is to say the crystal population undergoes a self-adaptive evolution process. So, it can effectively find materials' stable and metastable structures under certain conditions only provided the chemical information of the materials.~\cite{muse} More importantly, {\sc Muse} generates the random structures of the first generation with symmetry constraints, and then largely shortens the optimization time of the first generation and increase the diversity of crystal population. The random structures can be created according to the randomly chosen space group numbers from 2 to 230 and Wyckoff positions must be fit to the atom-number ratio of different kinds.~\cite{muse} Tests on systems including metallic, covalent and ionic systems all show that {\sc Muse} has very high efficiency and almost 100\% success rate.

The structures generated by the {\sc Muse} code were optimized by {\sc vasp} package.~\cite{Kresse94,Kresse96} We applied the generalized gradient approximation (GGA) parametrized by PBE~\cite{Perdew1996} and the electron-ion interaction described by the PAW scheme.~\cite{Blochl1994,Kresse99} The pseudopotential for Ta has the valence electrons' configuration of $\mathrm{5p^66s^25d^3}$. To achieve good convergences the kinetic energy cutoff and the \textit{k}-point grids spacing were chosen to be 500 eV and 0.03 $\mathrm{\AA}^{-1}$ in the calculations, respectively. The accuracies of the target pressure and the energy convergence for all optimizations are better than 0.1 GPa and $10^{-5}$ eV, respectively.  The systems containing 4, 8 and 16 atoms in the primitive cell were used in all the structure searches.

\section{Results and discussion}
\label{results}

\subsection{Structure search for Ta under high pressure}

\begin{figure}[!h]
\begin{center}
\includegraphics*[width=9.5cm]{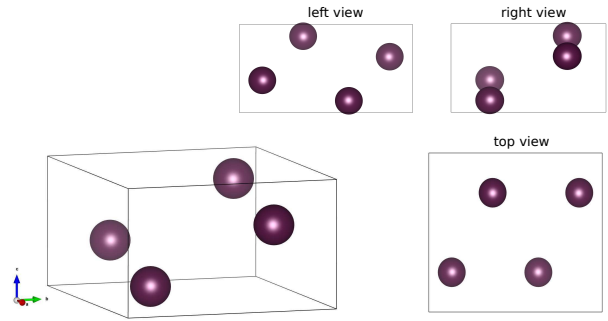}
\caption{(Color online) The crystal structure of \textit{Pnma}-Ta}
\label{fig:pnma-ta}
\end{center}
\end{figure}

The structures were generated with symmetry constraints and optimized at fixed pressure in our MAC crystal structure searches.~\cite{muse} The enthalpies of these structures were calculated and compared to find the proper path towards the lowest-enthalpy structure. Bcc has the lowest enthalpy among all the searched structures. It is interesting that we found an energetically competitive structure with the orthorhombic \textit{Pnma} symmetry (space group: 62). \textit{Pnma}-Ta has four atoms in its primitive cell (see Fig.~\ref{fig:pnma-ta}). The four atoms are at Wickoff's 4\textit{c} positions (0.132,  0.250,  0.366) with the lattice constants 4.98, 4.30 and 2.56 $\mathrm{\AA}$ at $\sim$100 GPa. This structure was then checked and confirmed by Burakovsky.~\cite{Burakovsky-pc} It is noted that we also reproduced the previous reported \textit{Pm$\bar3$n} (\textit{A}15) ~\cite{Streitz2002,Jakse2004} and \textit{P6/mmm} (hex-$\omega$)~\cite{Burakovsky2010,Hsiung1998-1,Hsiung2000,Hsiung2010} structures whose energy are much higher than \textit{Pnma} structure. Meanwhile, other produced structures of Ta with space groups \textit{I4/mmm}, \textit{C2/c}, \textit{Fddd} and so on, were also found to have much higher energies than \textit{Pnma}-Ta.

\begin{figure}[!h]
\begin{center}
\includegraphics*[width=6cm,angle=-90]{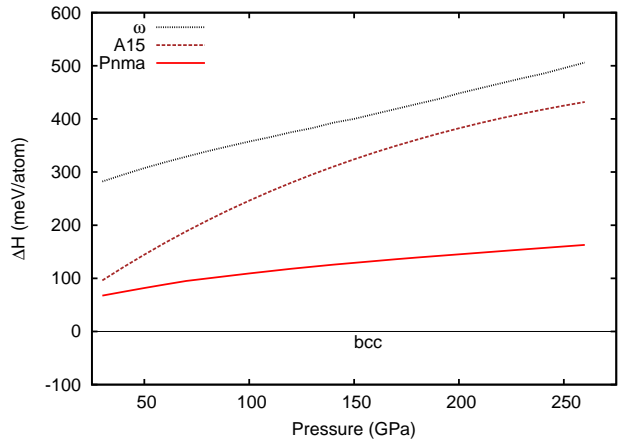}
\caption{(Color online) Enthalpy differences vs pressure with respective to bcc}
\label{fig:deltaH}
\end{center}
\end{figure}

In detail, we found that bcc is stable at pressure up to 300 GPa. The enthalpy differences of some low-enthalpy structures relative to bcc are plotted in Fig.~\ref{fig:deltaH}. We note that \textit{Pnma}-Ta has much lower enthalpies than the previously reported hex-$\omega$~\cite{Burakovsky2010,Hsiung1998-1,Hsiung2000,Hsiung2010} and \textit{A}15~\cite{Streitz2002,Jakse2004} phases. But it has lower symmetry than hex-$\omega$ (P6/mmm, 191) and \textit{A}15 (\textit{Pm$\bar3$n}, 223). Their enthalpy differences with respective to bcc are all positive and increase with pressure. This indicates that, at 0 K, bcc is the most stable phase and becomes more and more stable as pressure increases. While, it is probably not the case at elevated temperature and the \textit{Pnma} structure is expected to be more stable than bcc at HPHT. After all, \textit{Pnma}-Ta has much lower energy that hex-$\omega$ and \textit{A}15 structures at high pressure. Before we treat \textit{Pnma} structure as an alternative metastable phase of Ta, we should check its stability at high pressure.

\subsection{Stability of \textit{Pnma}-Ta}

\begin{figure}[!top]
\begin{center}
\includegraphics*[width=7.5cm]{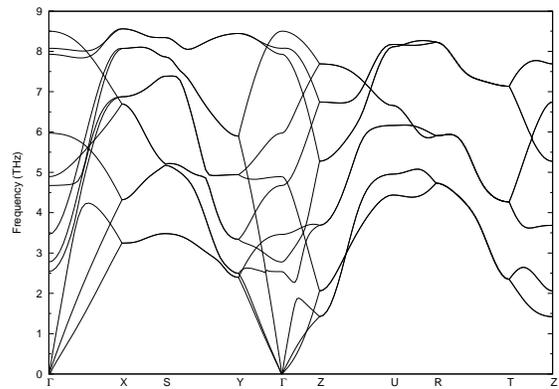}
\caption{Phonon dispersion curve of \textit{Pnma}-Ta at 96 GPa}
\label{fig:pnmaphon}
\end{center}
\end{figure}

To test the dynamic stability of this candidate metastable structure, we calculated its phonon dispersion curve. We determined the vibrational frequencies of \textit{Pnma}-Ta using the density functional perturbation theory (DFPT),~\cite{Baroni1987,Baroni2001} as implemented in the {\sc quantum-espresso} package.~\cite{pwscfcite} We used the generalized gradient approximation (GGA) proposed by Perdew, Burke and Ernzerhof (PBE)~\cite{Perdew1996} as the exchange-correlation functional. A nonlinear core correction to the exchange-correlation energy function was introduced to generate a Vanderbilt ultrasoft pseudopotential for Ta with the valence electrons’ configuration 4s$^2$5p$^6$5d$^3$6s$^2$, and the ultrasoft pseudopotentials were generated with a scalar-relativistic calculation. As an example, the calculated phonon dispersion curve of \textit{Pnma}-Ta at 96 GPa was shown in Fig.~\ref{fig:pnmaphon}. There are no imaginary frequencies in any position of the first Brillouin zone at pressure up to $\sim$250 GPa. This indicates the predicted \textit{Pnma}-Ta is dynamically stable under compression.

The mechanical stability of material is reflected by the elastic constants. We calculated the high-pressure elastic constants of \textit{Pnma}-Ta through stress-strain relationship, similar to Ref.~\onlinecite{Jochym2000}. All the resulting elastic constants of \textit{Pnma}-Ta are positive and increase with pressure (Table.~\ref{tab:pnmaec}). To check its mechanical stability under compression, we used the stability criteria of orthorhombic crystal,~\cite{Wu2007}

$C_{11} > 0,\quad C_{22} > 0,\quad C_{33} > 0,$

$C_{44} > 0,\quad C_{55} > 0,\quad C_{66} > 0,$

$C_{11} + C_{22} + C_{33} + 2(C_{12}+C_{13}+C_{23}) > 0,$

$C_{11} + C_{22} - 2C_{12} > 0,$

$C_{11} + C_{33} - 2C_{13} > 0,$

$C_{22} + C_{33} - 2C_{23} > 0.$\\
After careful check, we found the calculated elastic constants conform to these stability criteria at all pressures, suggesting \textit{Pnma}-Ta is also mechanically stable in the pressure range of interest.

\begin{table}[!top]
\caption{Elastic constants of \textit{Pnma}-Ta at different pressures $P$ and atomic volumes $V$. The elastic constants are all in GPa.}
\label{tab:pnmaec}
\begin{tabular}{rrrrrrrrrrr}
\hline\hline
$P$(GPa) & $V$($\mathrm{\AA^3}$) & $C_{11}$ & $C_{22}$ & $C_{33}$ & $C_{12}$ & $C_{13}$ & $C_{23}$ & $C_{44}$ & $C_{55}$ & $C_{66}$\\
\hline
41.1  & 15.75  & 513  & 535  & 492  & 249 & 190 & 235 & 99  & 86  & 132 \\  
52.5  & 15.25  & 575  & 600  & 568  & 269 & 206 & 248 & 118 & 98  & 151 \\
65.5  & 14.75  & 646  & 674  & 650  & 290 & 227 & 261 & 140 & 112 & 172 \\
80.5  & 14.25  & 727  & 759  & 746  & 316 & 250 & 276 & 165 & 128 & 196 \\
118.3 & 13.25  & 936  & 965  & 997  & 384 & 301 & 307 & 227 & 171 & 252 \\
142.1 & 12.75  & 1066 & 1087 & 1147 & 431 & 333 & 327 & 264 & 196 & 286 \\
170.0 & 12.25  & 1210 & 1226 & 1315 & 483 & 371 & 356 & 304 & 224 & 326 \\
203.0 & 11.75  & 1372 & 1385 & 1502 & 551 & 414 & 390 & 348 & 254 & 371 \\
242.2 & 11.25  & 1569 & 1584 & 1727 & 630 & 467 & 429 & 395 & 298 & 423 \\
\hline\hline
\end{tabular}
\end{table}

\subsection{Sound velocities of \textit{Pnma}-Ta}
\begin{figure}[!h]
\begin{center}
\includegraphics*[width=8cm]{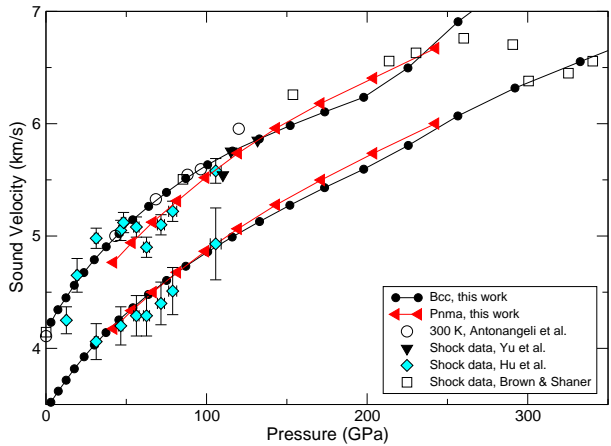}
\caption{(Color online) High pressure sound velocities of bcc and \textit{Pnma} Ta compared with experiments. The experimental data are from Refs.~\onlinecite{Brown1984,Antonangeli2010,Hu2012} and \onlinecite{Yu2006}.}
\label{fig:soundv}
\end{center}
\end{figure}

From the elastic constants, we deduced its high-pressure sound velocities in the way similar to our previous work.~\cite{Liu2011-2} To compare with experiments, we also calculated the elastic constants of bcc and deduced its high-pressure velocities with the same method. The comparison of the calculated sound velocities with experiments are plotted in Fig.~\ref{fig:soundv}. It is noted that our calculated sound velocities are at 0 K. So, one should be careful in the comparison. Below $\sim$100 GPa, our obtained bcc shear sound velocities are in excellent agreement with the results from the inelastic x-ray scattering (IXS) experiment at room temperature.~\cite{Antonangeli2010} While at 100 GPa, the shear sound velocity begins the softening behavior, consistent with the IXS experiment,~\cite{Antonangeli2010} and it recovers at about 250 GPa.

Recently, Hu \textit{et al.} observed a discontinuation of Hugoniot shear sound velocity $C_l$ at $\sim$60 GPa, implying a SS phase transition under shock compression.~\cite{Hu2012} Our calculated 0 K bcc-Ta sound velocities are in good agreement with their experimental data below 50 GPa. While, their experimental shear sound velocities have an abrupt decrease at about 60 GPa, implying a SS phase transition. Our bcc-Ta shear sound velocities have no such discontinuation at 60 GPa. But our shear sound velocities of \textit{Pnma}-Ta deviate from those of bcc-Ta. It is interesting that our calculated shear sound velocities of \textit{Pnma}-Ta reproduced the experimental discontinuation of shear sound velocities above 60 GPa. For bulk sound velocity $C_b$, our calculated values of both bcc and \textit{Pnma} are in good agreement with experiments, respectively, and they have no discontinuation in the pressure range of interest. Our shear sound velocities of \textit{Pnma}-Ta are also in good agreement with those of Hu \textit{et al.} This suggests that solid Ta transits from bcc to \textit{Pnma} phase under shock compression. It is noted that both our calculated 0 K shear sound velocities of bcc-Ta and the root temperature IXS data~\cite{Antonangeli2010} show the elastic softening at 100 GPa, but the Hugoniot shear sound velocities~\cite{Hu2012} have softening at 60 GPa. This is attributed to the effects of the high temperature and the shock shear stress, which lower the phase transition pressure.

\subsection{Stability of hex-$\omega$ Ta}


In order to determine the mechanical stability of hex-$\omega$-Ta, we also calculated its high-pressure elastic constants after full relaxation. For hexagonal lattices, there are five independent elastic constants, i.e. $C_{11}$, $C_{12}$, $C_{13}$, $C_{33}$ and $C_{44}$. We calculated the elastic constants as the second derivatives of the internal energy with respect to the strain tensor which leads to volume conserving lattice. The elastic constants were calculated at the equilibrium relaxed structure at fixed volume $V$ by keeping the strains in the lattice and relaxing the symmetry allowed internal degrees of freedom, and by finding the quadratic coefficients after fitting the forth-order polynomial curves to the total energies and strains. The method for reducing the five elastic constants is the same as that in Ref.~\onlinecite{Steinle-Neumann99}. 
%
%
%
%

\begin{figure}[!top]
\begin{center}
\includegraphics*[width=8.8cm]{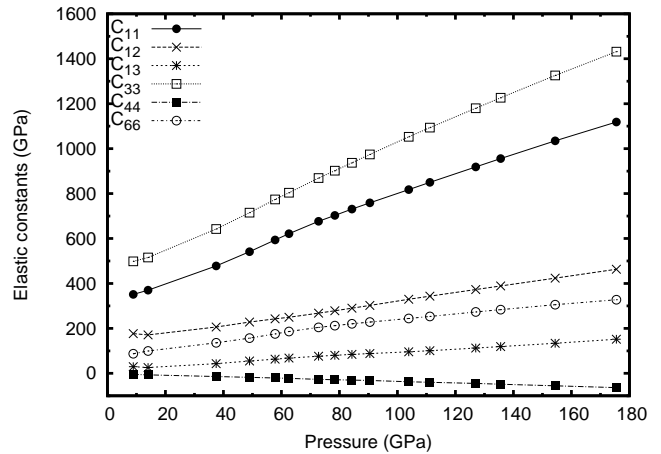}
\caption{The elastic constants of hex-$\omega$ Ta vs pressure.}
\label{fig:omelastic}
\end{center}
\end{figure}


The elastic constants of hex-$\omega$-Ta as the function of pressure are plotted in Fig.~\ref{fig:omelastic}. As one can see, all the elastic constants except $C_{44}$ are positive and increase with pressure. Shear modulus $C_{44}$ is negative and decreases with increasing pressure. This indicates that hex-$\omega$ phase is mechanically unstable at high pressure and 0 K. However, one can not conclude that it is unstable at high-temperature. So, we should check the high-temperature stability of hex-$\omega$ phase from the high-temperature phonon dispersion. 


It was pointed that the mechanically unstable phase at 0 K can be stabilized by vibrational entropy of lattice at elevated temperature, i.e. the interactions between phonons.~\cite{Souvatzis2008} The harmonic phonon calculation methods, typically the frozen phonon method and the density functional perturbation theory (DFPT),~\cite{Baroni1987,Baroni2001} only treat with the 0 K lattice, despite that they can include the electronic temperature through the finite-temperature DFT. In order to include the high-temperature lattice vibration, we applied the self-consistent \textit{ab initio} lattice dynamics (SCAILD) method recently developed by Souvatzis \textit{et al.}~\cite{Souvatzis2008} It has been successfully applied to many metals and other systems.~\cite{Souvatzis2008,Souvatzis2008-2,Luo2010,Souvatzis2009,Bozin2010,Souvatzis2010,Souvatzis2011,Souvatzis2011-2} The SCAILD method which is based on the calculation of Hellman-Feynman forces on atoms in a supercell is the robust extension of the frozen phonon method. It takes into account the anharmonic effects induced by the interactions between phonons.

In the SCAILD method, all phonons are excited together in the same cell by displacing atoms situated at the undistorted positions $\mathbf{R+b_{\sigma}}$, and the distorted positions are changed to $\mathbf{R+b_{\sigma}+U_{R\sigma}}$, where the displacements are written as~\cite{Souvatzis2008,Souvatzis2010}
\begin{equation}
	\mathbf{U_{R\sigma}}=\frac{1}{\sqrt{N}}\sum_{\mathbf{q,s}}\mathcal{A}_{\mathbf{qs}}^{\sigma}\epsilon_{\mathbf{qs}}^{\sigma}e^{i\mathbf{q(R+b_{\sigma})}}.
  \label{displ}
\end{equation}

In Eq.(\ref{displ}), $\mathbf{R}$ is the $N$ Bravais lattice sites of the supercell, $\mathbf{b_{\sigma}}$ the position of atom $\sigma$ with respect to this site, $\epsilon_{\mathbf{qs}}^{\sigma}$ are the phonon eigenvectors corresponding to the phonon mode, $s$. The mode amplitude $\mathcal{A}_{\mathbf{qs}}^{\sigma}$ can be evaluated from the different phonon frequencies $\omega_{\mathbf{q,s}}$ through~\cite{Souvatzis2008,Souvatzis2010}
\begin{equation}
\mathcal{A}_{\mathbf{qs}}^{\sigma}=\pm\sqrt{\frac{\hbar}{2M_{\sigma}\omega_{\mathbf{ks}}}coth\left(\frac{\hbar\omega_{\mathbf{qs}}}{2k_BT}\right)},
  \label{Aqs}
\end{equation}
where $T$ is the temperature of the system. The phonon frequencies~\cite{Souvatzis2008,Souvatzis2010}
\begin{equation} \omega_{\mathbf{qs}}=\left[\sum_{\sigma}\frac{\epsilon_{\mathbf{qs}}^{\sigma}\mathbf{F_{q}^{\sigma}}}{\mathcal{A}_{\mathbf{qs}}^{\sigma M_{\sigma}}}\right]^{1/2},
  \label{wqs}
\end{equation}
are obtained from the Fourier transform $\mathbf{F_{q}^{\sigma}}$ of the forces acting on the atoms in the supercell. The phonon frequencies are calculated in a self-consistent manner according to Eqs.\ref{displ}-\ref{wqs}, through which we alternate the forces on the displaced atoms and calculate new phonon frequencies and new displacements. The interaction between different lattice vibrations are included in SCAILD as the simultaneous presence of all the commensurate phonons in the same force calculation. The phonon frequencies are thus renormalized by the very same interaction.\cite{Souvatzis2010} 

\begin{figure}[!top]
\begin{center}
\includegraphics*[width=5cm,angle=-90]{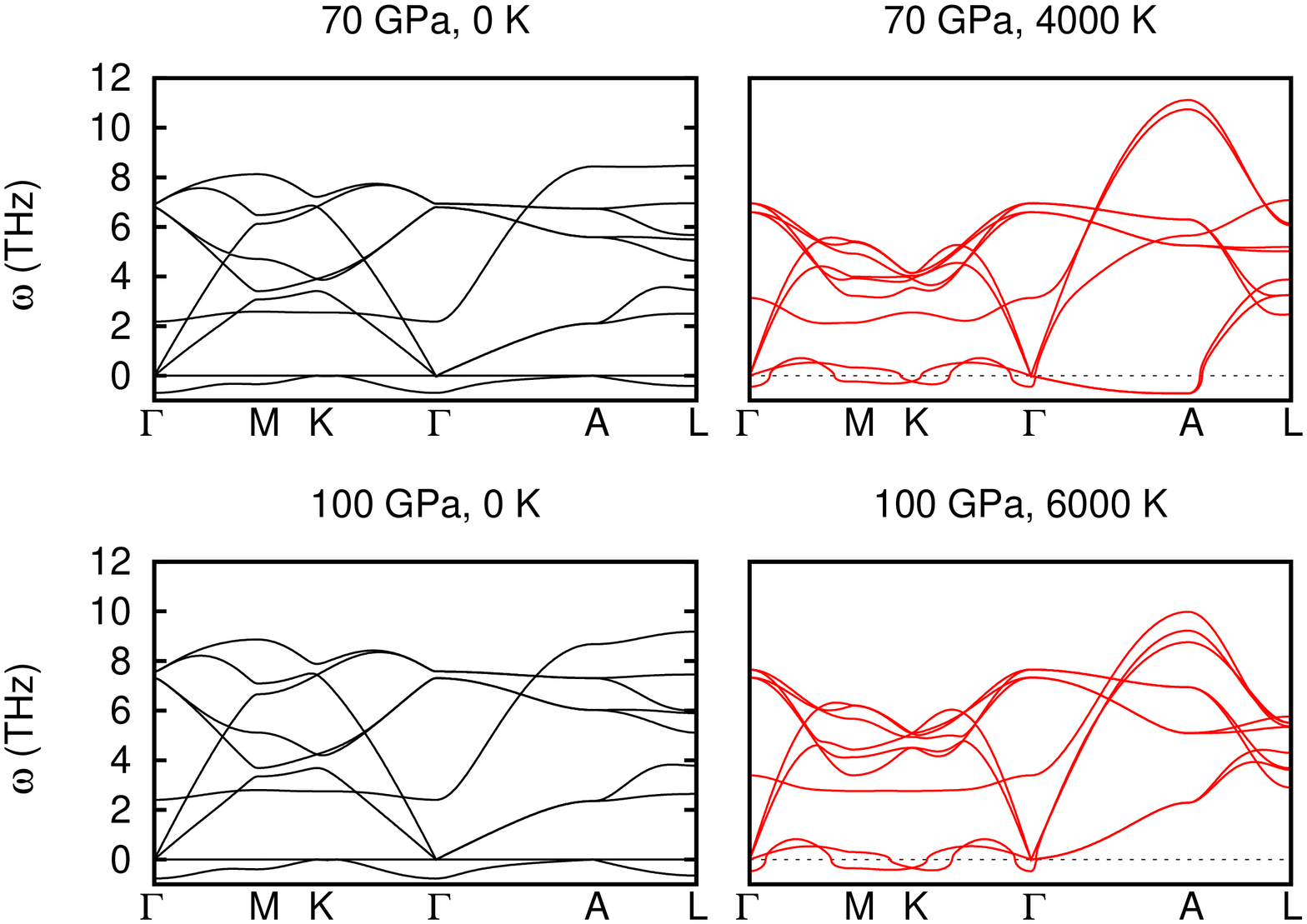}
\includegraphics*[width=5cm,angle=-90]{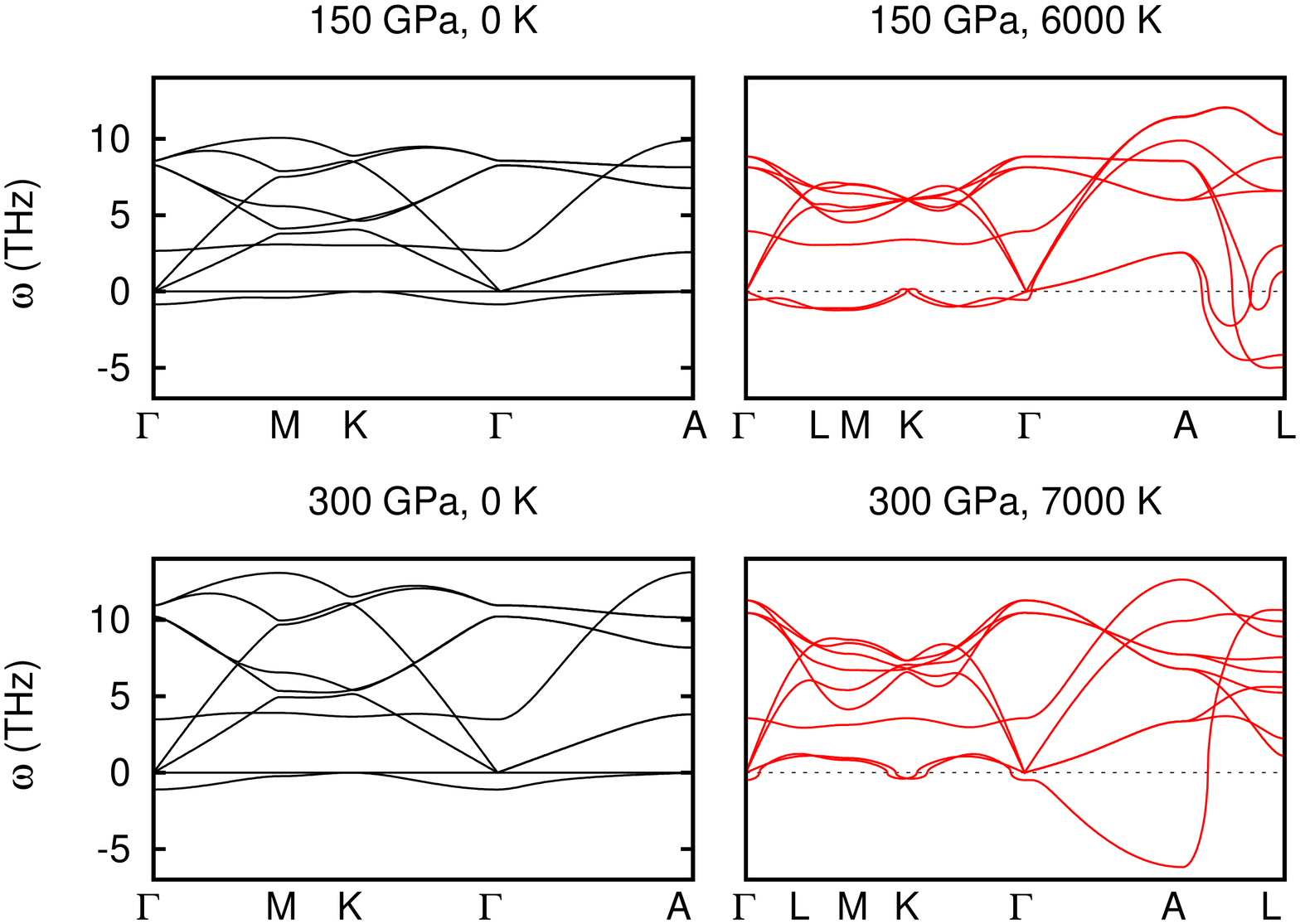}
\caption{(Color online) The 0 K and high-temperature phonon dispersion curves of hex-$\omega$-Ta at different pressures.}
\label{fig:Ta-omega}
\end{center}
\end{figure} 

The force calculations were performed with {\sc vasp},~\cite{Kresse94,Kresse96} within GGA parametrized by PBE.~\cite{Perdew1996} The electron-ion interactions were also described by the PAW scheme~\cite{Blochl1994,Kresse99} with the energy cutoff of 350 eV. Fermi-Dirac temperature smearing was applied to the Kohn-Sham occupational number together with a $6\times6\times6$ Monkhorst-Pack k-point grid. The electronic temperature was set as the same of lattice through Fermi smearing width. After tests, the supercell was chosen to contain 54 atoms resulted from the $3\times3\times2$ replication of hex-$\omega$ unit cell.

In the SCAILD calculations, the forces calculations are very demanding, and the high temperature phonon dispersion spectra were converged after more than 100 SCAILD loops. We calculated the finite temperature phonon dispersion from 0 to 300 GPa. As examples, the phonon dispersion spectra at 70, 100, 150 and 300 GPa are plotted in Fig.~\ref{fig:Ta-omega}. At 0 K, all the dispersion curves contain considerable imaginary frequencies in the acoustic branches, in agreement with the calculations of Burakovsky \textit{et al.}~\cite{Burakovsky2010} And at high temperature the converged phonon dispersion curves still have considerable imaginary frequencies at pressures from 0 to 300 GPa and temperatures from 4000 to 7000 K. So according to the SCAILD theory, hex-$\omega$ phase is unstable at high temperature under compression. This does not support the previous theoretical predictions that bcc-$\omega$ transition occurs before melting above 70 GPa.~\cite{Burakovsky2010}

\subsection{Discussion}
From our elastic constants and high-temperature phonon dispersion curves, we see that the ideal hex-$\omega$ phase is not stable at 0 K and elevated temperature under high pressure. Furthermore, the hex-$\omega$ phase is not energetically competitive compared with the \textit{Pnma} structure according to our structure searches. However, what Hsiung observed in the recent experiment is the pseudo-hex-$\omega$ with fractional coordinates: (0,0,0), (2.056/3,0.944/3,1/2), and (0.944/3,2.056/3,1/2).~\cite{Hsiung2010} We started from these initial coordinates in our optimizations, but obtained the ideal hex-$\omega$ with coordinates: (0,0,0), (2/3,1/3,1/2), and (1/3,2/3,1/2). And the high-temperature phonon dispersion curves of hex-$\omega$ Ta also show considerable imaginary frequencies, suggesting it is not dynamically stable. The negative shear moduli $C_{44}$ of hex-$\omega$ Ta reflected its mechanical instability. Why the pseudo-hex-$\omega$ structure is stable and observed in experiment is probably because of the existence of impurities such as W atoms~\cite{Hsiung2010} in the samples. W atoms prevent Ta atoms from moving to their ideal equilibrium positions in hexagonal lattice.

From the \textit{ab initio} molecular dynamics simulations of Ta, Wu \textit{et al.}~\cite{Wu2009} found that shear induces bcc to transit to a viscous plastic flow (partially disordered, partially crystalline structure) before melting. So they argued that this transition was misinterpreted as melting in previous DAC~\cite{Errandonea2001,Errandonea2003} experiments due to the similarity of the plastic flow to liquid. The substantial elastic softening and the discontinuities of the Hugoniot shear sound velocities~\cite{Hu2012} and the shear sound speed at room temperature~\cite{Burakovsky2010} of the bcc phase imply the SS phase transition in solid Ta. Just as Klug reviewed, possible additional phases except $\omega$ may complicate the measurement of the melting point and the pyrometric techniques also have space to improve.~\cite{Klug2010} The additional phases with very low symmetries resemble the local structure in the liquid phase. So, the transformations from bcc to such low symmetry phases bring the difficulties to identify the occurrence of melting in experiments. Our predicted new phase with \textit{Pnma} low symmetry is expected to be the phase to which bcc transits before melting.

\section{Conclusions}
\label{concl}
In conclusion, in order to find low-energy structures of Ta at high pressure, we globally searched the structures for pure Ta from 0 to 300 GPa using the \textit{ab initio} MAC crystal structure prediction technique. We found that bcc is the unique stable phase in this pressure range. It is interesting that we also found the energetically competitive \textit{Pnma}-Ta. \textit{Pnma}-Ta was tested to be more stable than hex-$\omega$ and \textit{A}15 structures. The calculated elastic constants of \textit{Pnma}-Ta satisfy the mechanical stability criteria of orthorhombic crystal. The absence of imaginary frequencies in \textit{Pnma}-Ta at high pressure indicates that it is also dynamically stable. To check the stability of the previously reported hex-$\omega$ Ta, we calculated its elastic constants and found it is not mechanically stable at high pressure and 0 K. To further explore the high temperature stability of hex-$\omega$ Ta, we calculated its high-temperature phonon spectra within the framework of the SCAILD approximation. The resulting high temperature phonon spectra contain considerable imaginary frequencies in the acoustic branches, also indicating the instability of hex-$\omega$ phase at HPHT. So we believe that Ta transits from bcc to \textit{Pnma} at HPHT before melting.


\section{acknowledgments}
The research was supported by the National Natural Science Foundation of China (11104127, 11104227 and U1230201/A06), the Project 2010A0101001 funded by CAEP, and the Science Research Scheme of Henan Education Department under Grand No. 2011A140019.

\bibliographystyle{apsrev4-1}
%
\end{document}